\documentclass[twocolumn,showkeys,preprintnumbers,amsmath,amssymb]{revtex4}

\usepackage{graphicx}
\usepackage{dcolumn}
\usepackage{bm}
\usepackage{bbm}

\begin{document}


\title{Convergence Analysis of the Wolf Method for Coulombic Interactions\footnote{To appear in \emph{Physics Letters A}.}}

\author{Arzhang Angoshtari}
\author{Arash Yavari}
 \email{arash.yavari@ce.gatech.edu}
\affiliation{ School of Civil and Environmental Engineering, Georgia
Institute of Technology, Atlanta, GA 30332. }

\date{\today}


\begin{abstract}
A rigorous proof for convergence of the Wolf method \cite{Wolf99} for calculating
electrostatic energy of a periodic lattice is presented. In
particular, we show that for an arbitrary lattice of unit cells, the
lattice sum obtained via Wolf method converges to the one obtained
via Ewald method.\end{abstract}


\maketitle

\section{Introduction}
The classical Madelung problem \citep{Madelung1918} has an important
role in atomic and molecular simulations involving electrostatic
interactions. Consider an arbitrary lattice with a unit cell that is
composed of $N$ charges $\{q_1,...,q_N\}$ and let linearly
independent vectors $\mathbf{e}_1,\mathbf{e}_2,\mathbf{e}_3 \in
\mathbb{R}^3$ denote the lattice vectors. We assume the charge
neutrality condition for the unit cell, i.e., $\sum_{i=1}^{N}q_i=0$.
Then the Madelung problem for calculation of the total electrostatic
energy of the unit cell located at the origin can be expressed as
\begin{equation}\label{EcellOr}
    \mathcal{E}_{\text{cell}}=\frac{1}{2}\sum_{i,j=1}^{N}q_{i}q_{j}\sideset{}{'}\sum_{\mathbf{n} \in
    \mathbb{Z}^3}
    |\mathbf{V}\mathbf{n}+\mathbf{r}_{ij} |^{-1},
\end{equation}
where $\mathbf{V}=[\mathbf{e}_1~\mathbf{e}_2~\mathbf{e}_3] \in
\mathbb{R}^{3\times3}$ (the matrix with lattice vectors as its
columns) and $\mathbf{r}_{ij}=\mathbf{r}_{i}-\mathbf{r}_{j}$, where
$\mathbf{r}_{i}$ denotes the atomic position within the unit cell.
The prime on the summation emphasizes that we exclude self-energy,
i.e., for $\mathbf{n}=\mathbf{0}$ the term $i=j$ is omitted. In
order to be able to use the well-established theory of
multi-dimensional zeta functions \citep{El98}, we introduce the
following non-standard representation of electrostatic energy of a
unit cell in an arbitrary lattice:
\begin{equation}\label{Ecell}
    \mathcal{E}_{\text{cell}}=\frac{1}{2}\sum_{i,j=1}^{N}q_{i}q_{j}\sideset{}{'}\sum_{\mathbf{n} \in
    \mathbb{Z}^3}
    \left[ \frac{1}{2}\left( \mathbf{n}+\mathbf{p}_{ij} \right)^{\textsf{T}}\mathbf{Q}\left( \mathbf{n}+\mathbf{p}_{ij} \right) \right]^{-1/2},
\end{equation}
where $\textsf{T}$ stands for matrix transport and
$Q_{ij}=2\mathbf{e}_i\cdot\mathbf{e}_j$ is a positive-definite
matrix (twice the metric tensor),
$\mathbf{p}_{ij}=\mathbf{U}\mathbf{r}_{ij}$ with
$\mathbf{U}=\mathbf{V}^{-1}$ and the prime on the summation denotes
the exclusion of the self energy. For example, for an orthorhombic
lattice with lattice parameters $a=|\mathbf{e}_{1}|$,
$b=|\mathbf{e}_{2}|$ and $c=|\mathbf{e}_{3}|$ we have
\begin{eqnarray}
  && \mathbf{Q}=\left(%
\begin{array}{ccc}
  2a^2 & 0 & 0 \\
  0 & 2b^2 & 0 \\
  0 & 0 & 2c^2 \\
\end{array}%
\right),~~
\mathbf{U}=\left(%
\begin{array}{ccc}
  \frac{1}{a} & 0 & 0 \\
  0 & \frac{1}{b} & 0 \\
  0 & 0 & \frac{1}{c
  } \\
\end{array}%
\right).\nonumber \\
\end{eqnarray}
For a hexagonal lattice with unit cell vectors
\begin{eqnarray}
  && \mathbf{e}_{1}=\left(%
\begin{array}{c}
  a \\
  0 \\
  0 \\
\end{array}%
\right),~~
    \mathbf{e}_{2}=\left(%
\begin{array}{c}
  \frac{1}{2}a \\
  \frac{\sqrt{3}}{2}a \\
  0 \\
\end{array}%
\right),~~
\mathbf{e}_{3}=\left(%
\begin{array}{c}
  0 \\
  0 \\
  c \\
\end{array}%
\right),  \nonumber \\
\end{eqnarray}
where $a$ and $c$ are unit cell parameters, we have
\begin{eqnarray}
  && \mathbf{Q}=\left(%
\begin{array}{ccc}
  2a^2 & a^2 & 0 \\
  a^2 & 2a^2 & 0 \\
  0 & 0 & 2c^2 \\
\end{array}%
\right),~~
\mathbf{U}=\left(%
\begin{array}{ccc}
  \frac{1}{a} & -\frac{1}{\sqrt{3}a} & 0 \\
  0 & \frac{2}{\sqrt{3}a} & 0 \\
  0 & 0 & \frac{1}{c} \\
\end{array}%
\right). \nonumber \\
\end{eqnarray}

One method of calculating the above lattice sum is to use direct
sums. However, it is a well known fact that (\ref{Ecell}) is a
conditionally convergent series, which means that (\ref{Ecell}) is
meaningless unless the order of summation of the terms is specified.
It is interesting to note that summation over regions that may seem
to be natural can diverge. As an example, for a $3$-dimensional NaCl-type
ionic crystal, it was shown that this lattice sum does not converge
over expanding spheres \citep{Bor85}, expanding ellipsoids, and some
specific expanding polygons \citep{Bor98}, but it would converge for
expanding cubes \citep{Bor85}. The direct summation method is not
practical due to the slow rate of convergence.

Another method for dealing with (\ref{Ecell}) is to find some
analytic continuation for this expression over the complex plane
and then to find some fast converging series to evaluate this
analytic continuation. The celebrated Ewald method \citep{Ewald21}
uses this procedure. One can write $\mathcal{E}_{\text{cell}}$ as
\begin{equation}
    \mathcal{E}_{\text{cell}}=\frac{1}{2}\sum_{i,j=1}^{N}q_{i}q_{j}Z_{\mathbf{Q}}(1,\mathbf{p}_{ij}),
\end{equation}
with
\begin{equation}\label{Z}
    Z_{\mathbf{Q}}(s,\mathbf{p})=\sideset{}{'}\sum_{\mathbf{n} \in \mathbb{Z}^3}
    \left[ \frac{1}{2}\left( \mathbf{n}+\mathbf{p} \right)^{\textsf{T}}\mathbf{Q}\left( \mathbf{n}+\mathbf{p} \right) \right]^{-s/2},
\end{equation}
where again the prime denotes the exclusion of any infinite summands.
$Z_{\mathbf{Q}}(s,\mathbf{p})$ is a special case of the general
Epstein zeta functions \citep{El98,Terras85}. It is known that
$Z_{\mathbf{Q}}(s,\mathbf{p})$ is uniformly and absolutely
convergent for any complex number $s$ with $\mathfrak{Re}(s)>3$ and
it has a meromorphic continuation to the whole complex s-plane
\citep{Terras85,El98,CranBuh87}. From here on, by
$Z_{\mathbf{Q}}(s,\mathbf{p})$ we mean its analytic continuation.
$Z_{\mathbf{Q}}(s,\mathbf{p})$ is analytic everywhere except for the
simple pole at $s=3$ with the residue \citep{El98}
\begin{equation}\label{ResZ_s3}
    \underset{s=3}{\mathrm{Res}}~Z_{\mathbf{Q}}(s,\mathbf{p})=\frac{(2\pi)^{3/2}}{\Gamma(3/2)\sqrt{\mathrm{det}\mathbf{Q}}}=\frac{2^{5/2}\pi}{\sqrt{\mathrm{det}\mathbf{Q}}}.
\end{equation}
Using some special functions, it is possible to write
$Z_{\mathbf{Q}}(s,\mathbf{p})$ and thus $\mathcal{E}_{\text{cell}}$
in terms of a rapid converging series that recovers the standard
Ewald method \citep{Fu94,Terras85}.

Determining the relation between the above-mentioned two methods is not a
straightforward task. Specifically, in what order one should sum the
terms of the series (\ref{Ecell}) to obtain the Ewald's result?
Borwein, et al. \citep{Bor85} showed that for NaCl-type crystals,
summing over cubes would yield the Ewald result. It is interesting
to note that $\sum_{\mathbf{n} \in
\mathbb{Z}^3}'|\mathbf{n}+\mathbf{r}|^{-1}$ is not a convergent
series and since all of the summands are positive, partial sums of
this series would be unbounded. Thus, it is meaningless to expect to
find $Z_{\mathbf{Q}}(1,\mathbf{p})$ using direct sums.

As we mentioned earlier, for NaCl-type crystals direct summation
over expanding cubes converges while summation over expanding
spheres diverges. One may guess that what makes the expanding cubes
to converge is that, unlike expanding spheres, each cube is charge
neutral, and thus it may be possible to obtain a converging sequence
over spheres if one somehow converts the regular spheres to charge
neutral ones. This is the main idea of the Wolf method
\citep{Wolf99} and the earlier work of Buhler and Crandall
\citep{BuhCran90}. In particular, for a general lattice of charges,
Wolf, et al. \citep{Wolf99} suggested that putting a mirror charge
on the surface of sphere for each charge inside the sphere and
neglecting the charges outside the sphere results in a convergent
sequence that converges to the result obtained via Ewald method.
Although they verified their method by considering several numerical
examples, they did not present a rigorous proof.

In this paper, we present the missing proof of the convergence of
Wolf's method for calculating electrostatic energy of an arbitrary
lattice of charges. We should mention that Buhler and Crandall
\citep{BuhCran90} presented a proof for NaCl lattice with unit
charges. Here we generalize their proof to arbitrary lattices. Note
that Wolf, et al. \citep{Wolf99} presented two differrent methods,
namely, damped and undamped methods. Here by Wolf method we mean the
undamped method. In \S 2 we present the proof and in \S 3 we mention
some concluding remarks and future directions.

\section {\textbf{Proof of the Convergence of Wolf's Method}}


We use the contour integral method of \citep{BuhCran90} to prove the
convergence of Wolf's method. Let us first review the required
preliminaries. We consider the following analytic continuation of
(\ref{Ecell})
\begin{eqnarray}\label{Es}
    \!\!\!&{}&\!\!\! E(\mathbf{Q},s)  \nonumber \\
    \!\!&{}&\!\! =\frac{1}{2}\sum_{i,j=1}^{N}q_{i}q_{j}\sideset{}{'}\sum_{\mathbf{n} \in \mathbb{Z}^3}
    \left[ \frac{1}{2}\left( \mathbf{n}+\mathbf{p}_{ij} \right)^{\textsf{T}}\mathbf{Q}\left( \mathbf{n}+\mathbf{p}_{ij} \right)
    \right]^{-s/2} \nonumber \\
    \!\!&{}&\!\! =\frac{1}{2}\sum_{i,j=1}^{N}q_{i}q_{j}Z_{\mathbf{Q}}(s,\mathbf{p}_{ij}).
\end{eqnarray}
Note that
$E(\mathbf{Q},1)=\mathcal{E}_{\text{Ewald}}=\mathcal{E}_{\text{cell}}$.
Let $\Psi(x)$ denote the function
\begin{equation}\
    \Psi(x)= \left\{ %
\begin{array}{c}
  0~~~~~~~x<1, \\
  \frac{1}{2} ~~~~~~~x=1,\\
  1~~~~~~~x>1. \\
\end{array}%
\right.\label{H}
\end{equation}
Then, for $R \in \mathbb{R}^{+}$, where $\mathbb{R}^{+}$ is the set
of the positive real numbers, define the spherically truncated
(finite) sum
\begin{eqnarray}\label{ERs}
    \!\!\! &{}& \!\!\! E_{R}(\mathbf{Q},s) = \nonumber \\
     \!\! &{}& \!\! \frac{1}{2} \sum_{i,j=1}^{N}q_{i}q_{j}\sideset{}{'}\sum_{\mathbf{n} \in
      \mathbb{Z}^3} \Bigg \{
     \Psi\left(\frac{R}{\left[ \frac{1}{2}\left( \mathbf{n}+\mathbf{p}_{ij} \right)^{\textsf{T}}\mathbf{Q}\left( \mathbf{n}+\mathbf{p}_{ij} \right)
    \right]^{1/2}}\right) \nonumber \\
     \!\! &{}& \!\! \times
      \left[ \frac{1}{2}\left( \mathbf{n}+\mathbf{p}_{ij} \right)^{\textsf{T}}\mathbf{Q}
     \left( \mathbf{n}+\mathbf{p}_{ij} \right) \right]^{-s/2}  \Bigg \}.
\end{eqnarray}
Note that $E_{R}(\mathbf{Q},s)$ roughly denotes the direct sum of
(\ref{Ecell}) over spheres centered at charges in the unit cell
located at the origin with radius $R$. From (\ref{H}) one can see
that if
\begin{equation}
    R=\mathcal{R}(\mathbf{n},\mathbf{p}_{ij}):=\left[ \frac{1}{2}\left( \mathbf{n}+\mathbf{p}_{ij}
\right)^{\textsf{T}}\mathbf{Q}\left( \mathbf{n}+\mathbf{p}_{ij}
\right) \right]^{1/2},
\end{equation}
for some values of $i$, $j$ and $\mathbf{n}$, then the charges
located on the surface of the sphere would have the additional
weight of $1/2$. But this would be irrelevant in our analysis
because as we will see in the sequel, we are interested in the
behavior of a continues function of $R$ as $R\rightarrow\infty$.
This can be expressed in term of $E_{R}(\mathbf{Q},s)$ if $R$ does
not take the discrete values
$\mathcal{R}(\mathbf{n},\mathbf{p}_{ij})$ for $\mathbf{n} \in
\mathbb{Z}^3$ and $i,j=1,\ldots,N$, and so we can exclude those
values.

In the Wolf method one considers spheres with radii $R$ centered at
each charge of the unit cell at the origin and puts a mirror charge
on the surface of the sphere for each charge inside the sphere
including center charges. Then energy of the unit cell,
$\mathcal{E}_{\text{Wolf}}$, is calculated using only charges
located on and inside these spheres. This can be written as
\begin{equation}\label{Ewolf1}
    \mathcal{E}_{\text{Wolf}}= \lim_{R\rightarrow\infty} \left[ E_{R}(\mathbf{Q},1)-\frac{1}{R}E_{R}(\mathbf{Q},0)-\frac{1}{R}\sum_{i=1}^{N}q_{i}^{2}\right].
\end{equation}
But for a lattice with $N$ charges in its unit cell, the term
$\sum_{i=1}^{N}q_{i}^{2}$ is bounded, and thus we have
\begin{equation}\label{Ewolf2}
    \mathcal{E}_{\text{Wolf}}= \lim_{R\rightarrow\infty} \left[ E_{R}(\mathbf{Q},1)-\frac{1}{R}E_{R}(\mathbf{Q},0)\right].
\end{equation}
For $c \in \mathbb{R}^{+}$, Perron's formula states that
\citep{Ti86}
\begin{equation}\label{Perron}
    \frac{1}{2\pi \mathbbm{i}}\int_{c-\mathbbm{i}\infty}^{c+\mathbbm{i}\infty}\frac{R^s}{s}ds=\Psi(R).
\end{equation}

\begin{figure}[t]
\begin{center}
\includegraphics[scale=0.7,angle=0]{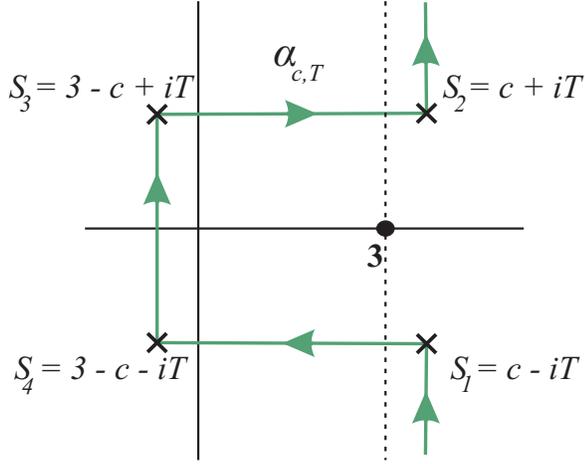}
\end{center}
\vspace*{-0.2in}\caption{\footnotesize The contour $\alpha_{c,T}$ in
the complex s-plane.} \label{fig1}
\end{figure}
Finally, for any $c,T\in \mathbb{R}^{+}$ with $c>3$ let
$\alpha_{c,T}$ be the contour in the complex s-plane depicted in
Fig. \ref{fig1}. Also let
$\mathcal{S}_{T}=\left\{s\in\mathbb{C}|~0<\mathfrak{Re}(s)<3
~\textrm{and}~|\mathfrak{Im}(s)|<T\right\}$. Then, the following
lemma holds.

\vskip 0.1 in \textbf{Lemma.} \textit{Let $\epsilon \in
\mathbb{R}^{+}$. Then there exist $c,T\in \mathbb{R}^{+}$ with
$c>3$, such that for any $w \in \mathcal{S}_{T}$ the following
relation holds as $R\rightarrow\infty$:
\begin{equation}\label{lemma}
    I_{\alpha}=\frac{1}{2\pi \mathbbm{i}}\int_{\alpha_{c,T}}\frac{Z_{\mathbf{Q}}(s,\mathbf{p})R^s}{s(s-w)}ds=O(R^{\epsilon}).
\end{equation}
}\vskip 0.1 in \textbf{Proof:} Let $s=\sigma+\mathbbm{i}t$ and
$I_{\alpha}=\sum_{i=1}^{5}I_i$, where
\begin{eqnarray}
&& I_{1}=\frac{1}{2\pi
\mathbbm{i}}\int_{c-\mathbbm{i}\infty}^{c-\mathbbm{i}T}\frac{Z_{\mathbf{Q}}(s,\mathbf{p})R^s}{s(s-w)}ds,\nonumber \\
&& I_{2}=\frac{1}{2\pi
    \mathbbm{i}}\int_{c-\mathbbm{i}T}^{3-c-\mathbbm{i}T}\frac{Z_{\mathbf{Q}}(s,\mathbf{p})R^s}{s(s-w)}ds,
    \nonumber \\
    && I_{3}=\frac{1}{2\pi \mathbbm{i}}\int_{3-c-\mathbbm{i}T}^{3-c+\mathbbm{i}T}\frac{Z_{\mathbf{Q}}(s,\mathbf{p})R^s}{s(s-w)}ds,\nonumber \\
    && I_{4}=\frac{1}{2\pi
    \mathbbm{i}}\int_{3-c+\mathbbm{i}T}^{c+\mathbbm{i}T}\frac{Z_{\mathbf{Q}}(s,\mathbf{p})R^s}{s(s-w)}ds,
    \nonumber \\
    && I_{5}=\frac{1}{2\pi
    \mathbbm{i}}\int_{c+\mathbbm{i}T}^{c+\mathbbm{i}\infty}\frac{Z_{\mathbf{Q}}(s,\mathbf{p})R^s}{s(s-w)}ds.
\end{eqnarray}
We are going to show that all of the above integrals are bounded.
For $I_1$ and $I_5$ we have $|Z_{\mathbf{Q}}(s,\mathbf{p})| =
|Z_{\mathbf{Q}}(c,\mathbf{p})|$, $|R^{s}| = R^c$ and
$[s(s-w)]^{-1}=O(t^{-2})$. Thus, these integrals are
$O(R^{c}T^{-1})$.

Using the standard methods of analytic number theory and the
Phragm\'{e}n-Lindel\"{o}f theorem \citep{Ivic85,Ti86}, we find that
$|Z_{\mathbf{Q}}(s,\mathbf{p})|=O(t^{c-3/2})$ for $3-c<\sigma<c$,
and hence $I_2$ and $I_4$ are $O(R^{c}T^{c-7/2})$. If $3<c<7/2$ and
$T=R^{7/(7-2c)}$, we can conclude that $I_1, I_2, I_4$ and $I_5$
are $O(R^\epsilon)$ for any $\epsilon>0$.

Next, note that \citep{El98}
\begin{eqnarray}\label{Zzeta}
    Z_{\mathbf{Q}}(s,\mathbf{p})&=&\sideset{}{'}\sum_{\mathbf{n} \in \mathbb{Z}^3}
    \left[ \frac{1}{2}\left( \mathbf{n}+\mathbf{p} \right)^{\textsf{T}}\mathbf{Q}\left( \mathbf{n}+\mathbf{p} \right)
    \right]^{-s/2} \nonumber \\
    &=&
    \sideset{}{'}\sum_{\mathbf{n} \in
    \mathbb{Z}^3}|\mathbf{A}\mathbf{n}-\mathbf{d}|^{-s}=:\zeta_{\mathbf{A}}(s,\mathbf{d}),
\end{eqnarray}
where $\mathbf{A}$ is a matrix with nonzero determinant. Define
\begin{equation}\label{Lam}
    \Lambda_{\mathbf{A}}(s,\mathbf{d})=\sqrt{\mathrm{det}\mathbf{A}}~\pi^{-s/2}~\Gamma\left(\frac{s}{2}\right)\zeta_{\mathbf{A}}(s,\mathbf{d}).
\end{equation}
Then, we have the following functional equation \citep{CranBuh87}
\begin{equation}\label{Lam2}
    \Lambda_{\mathbf{A}}(s,\mathbf{d})=e^{\pi |\mathbf{d}|^{2}
    \mathbbm{i}}\Lambda_{\mathbf{B}}(3-s,\mathbf{0}),
\end{equation}
where $\mathbf{B}=\mathbf{A}^{-\textsf{T}}$ with $-\textsf{T}$
denoting the inverse transpose. Substituting (\ref{Lam}) into
(\ref{Lam2}) yields
\begin{eqnarray}\label{zeta2}
    \zeta_{\mathbf{A}}(s,\mathbf{d})=\frac{e^{\pi |\mathbf{d}|^{2}
    \mathbbm{i}}~\pi^{s-3/2}~\Gamma\left[ \frac{1}{2}(3-s) \right]}{\mathrm{det}\mathbf{A}~\Gamma\left(\frac{s}{2}\right)}~\zeta_{\mathbf{B}}(3-s,\mathbf{0
    }).\nonumber \\
\end{eqnarray}
Now we replace $s$ by $3-s$ in $I_{3}$ and use (\ref{Zzeta}) and
(\ref{zeta2}) to write the integral over the upper half segment as
\begin{eqnarray}\label{I3t}
\!\!\! &{}& \!\!\!\frac{1}{2\pi
\mathbbm{i}}\int_{3-c}^{3-c+\mathbbm{i}T}\frac{Z_{\mathbf{Q}}(s,\mathbf{p})R^s}{s(s-w)}ds=\nonumber
\\ \!\!\!\!\! &{}& \!\!\!\!\! \frac{1}{2\pi
\mathbbm{i}}\int_{c}^{c+\mathbbm{i}T}\frac{Z_{\mathbf{Q}}(3-s,\mathbf{p})R^{3-s}}{(s-3)(s+w-3)}ds=
\nonumber \\ \!\!\!\!\! &{}& \!\!\!\!\! \frac{1}{2\pi
\mathbbm{i}}\int_{c}^{c+\mathbbm{i}T}\frac{\zeta_{\mathbf{A}}(3-s,\mathbf{d})R^{3-s}}{(s-3)(s+w-3)}ds=
\nonumber \\ \!\!\!\!\! &{}& \!\!\!\!\! \frac{ e^{\pi
|\mathbf{d}|^{2}\mathbbm{i}}~ \pi^{1/2}  }
{2(\mathrm{det}\mathbf{A})\mathbbm{i}}\int_{c}^{c+\mathbbm{i}T}
 \frac{\pi^{-s}~\Gamma\left(\frac{s}{2}\right) \zeta_{\mathbf{B}}(s,\mathbf{0}) R^{3-s}~ ds} {\Gamma\left[ \frac{1}{2}(3-s) \right]
 (s-3)(s+w-3)}=
 \nonumber \\ \!\!\!\!\! &{}& \!\!\!\!\! \frac{ e^{\pi |\mathbf{d}|^{2}\mathbbm{i}}~ \pi^{1/2}  }
{2(\mathrm{det}\mathbf{A})\mathbbm{i}} \sideset{}{'}\sum_{\mathbf{n}
\in
    \mathbb{Z}^3}  \int_{c}^{c+\mathbbm{i}T}
 \frac{\pi^{-s}~\Gamma\left(\frac{s}{2}\right) |\mathbf{A}\mathbf{n}|^{-s} R^{3-s}~ds } {\Gamma\left[ \frac{1}{2}(3-s) \right]
 (s-3)(s+w-3)},\nonumber \\
\end{eqnarray}
where in the last step, since $c>3$, we use the fact that
$\zeta_{\mathbf{B}}(s,\mathbf{0})$ is uniformly and absolutely
convergent over the integration path. Stirling's formula states that
for any fixed strip $\alpha\leq\sigma\leq\beta$, as
$t\rightarrow\infty$ \citep{Ti86}
\begin{eqnarray}\label{Stirling}
    \!\!\! &{}& \!\!\! \log\left[ \Gamma(\sigma+\mathbbm{i}t)\right]=\nonumber \\
    \!\!\! &{}& \!\!\!
    \left(\sigma+\mathbbm{i}t-\frac{1}{2}\right)\log(\mathbbm{i}t)-\mathbbm{i}t+\frac{1}{2}\log(2\pi)+O(t^{-1}). \nonumber \\
\end{eqnarray}
Using (\ref{Stirling}), we conclude that to bound (\ref{I3t}) one
needs to bound
\begin{equation}\label{intg}
    R^{3-c}\sum_{\mathbf{n} \in
    \mathbb{Z}^3}|\mathbf{A}\mathbf{n}|^{-c}\int_{t_0}^{T}t^{c-7/2} e^{\mathbbm{i}\left[t\log(t)-t-t\log(\pi R|\mathbf{A}\mathbf{n}|)
    \right]}ds,
\end{equation}
where $t_{0} >0$ is an arbitrary constant. But using the method of
stationary phase \citep{Murr84}, it can be shown that this integral
is $O(\log R)$ \citep{BuhCran90} and so $I_3$ is $O(R^\epsilon)$ for
$\epsilon>0$. To summarize, we have proved that $I_{1},\ldots,I_{5}$
are $O(R^\epsilon)$ for $\epsilon>0$ and hence (\ref{lemma}) holds.
$~~~~~\square$

Now we state the main result of this paper.

\vskip 0.1 in \textbf{Theorem.} \textit{There exists $T\in
\mathbb{R}^{+}$ such that if $s \in \mathcal{S}_{T}$ then
\begin{equation}\label{Es_theorem}
    E(\mathbf{Q},s)= \lim_{R\rightarrow\infty} \left[ E_{R}(\mathbf{Q},s)-\frac{1}{R^s}E_{R}(\mathbf{Q},0)\right].
\end{equation}
In particular, setting $s=1$ yields
\begin{eqnarray}\label{Es_theorem2}
    \mathcal{E}_{\text{cell}}&=&E(\mathbf{Q},1)= \lim_{R\rightarrow\infty} \left[ E_{R}(\mathbf{Q},1)-\frac{1}{R}E_{R}(\mathbf{Q},0)\right]\nonumber\\
    &=&\mathcal{E}_{\text{Wolf}}.
\end{eqnarray}
}\vskip 0.1 in \textbf{Proof:} Choose $c,T\in \mathbb{R}^{+}$ in
accordance with the previous lemma and let $w \in \mathcal{S}_{T}$.
Consider the integral
\begin{eqnarray}\label{intFws1}
    I&=&\frac{1}{2\pi \mathbbm{i}}\int_{c-\mathbbm{i}\infty}^{c+\mathbbm{i}\infty}\frac{E(\mathbf{Q},s)R^{s}}{s-w}ds\nonumber\\
    &=&\frac{1}{2\pi \mathbbm{i}}
    \int_{c-\mathfrak{Re}(w)-\mathbbm{i}\infty}^{c-\mathfrak{Re}(w)+\mathbbm{i}\infty}
    \frac{E(\mathbf{Q},s+w)R^{s+w}}{s}ds.
\end{eqnarray}
Note that since $c>3$, the contour of the integral is in the region
of the uniform and absolute convergence of (\ref{Ecell}). Also
$c-\mathfrak{Re}(w)>0$, and hence using (\ref{Perron}) one can write
(\ref{intFws1}) as
\begin{eqnarray}\label{intFws2}
    \!\!\! &{}& \!\!\!I=\frac{1}{2\pi \mathbbm{i}}\int_{c-\mathfrak{Re}(w)-\mathbbm{i}\infty}^{c-\mathfrak{Re}(w)+\mathbbm{i}\infty}
     \Bigg\{ \frac{R^{s+w}}{s} \times \nonumber\\
    \!\!\! &{}& \!\!\!  \frac{1}{2}\sum_{i,j=1}^{N}q_{i}q_{j}\sideset{}{'}\sum_{\mathbf{n} \in \mathbb{Z}^3}
    \left[ \frac{1}{2} \left( \mathbf{n}+\mathbf{p}_{ij} \right)^{\textsf{T}}\mathbf{Q}\left( \mathbf{n}+\mathbf{p}_{ij} \right)
    \right]^{-(s+w)/2} \Bigg \}ds \nonumber\\
    \!\!\! &{}& \!\!\! = \frac{R^w}{2}\sum_{i,j=1}^{N}q_{i}q_{j}\sideset{}{'}\sum_{\mathbf{n} \in \mathbb{Z}^3}
    \left[ \frac{1}{2}\left( \mathbf{n}+\mathbf{p}_{ij} \right)^{\textsf{T}}\mathbf{Q}\left( \mathbf{n}+\mathbf{p}_{ij} \right) \right]^{-w/2}\times \nonumber\\
      \!\!\! &{}& \!\!\!\frac{1}{2\pi \mathbbm{i}}\int_{c-\mathfrak{Re}(w)-\mathbbm{i}\infty}^{c-\mathfrak{Re}(w)+\mathbbm{i}\infty}
    \left( \frac{R}{\left[ \frac{1}{2}\left( \mathbf{n}+\mathbf{p}_{ij} \right)^{\textsf{T}}\mathbf{Q}\left( \mathbf{n}+\mathbf{p}_{ij} \right) \right]^{1/2}} \right)^{s}
    \frac{ds}{s} \nonumber\\
    \!\!\! &{}& \!\!\! = \frac{R^w}{2}\sum_{i,j=1}^{N}q_{i}q_{j}\sideset{}{'}\sum_{\mathbf{n} \in \mathbb{Z}^3}
    \left[ \frac{1}{2}\left( \mathbf{n}+\mathbf{p}_{ij} \right)^{\textsf{T}}\mathbf{Q}\left( \mathbf{n}+\mathbf{p}_{ij} \right) \right]^{-w/2} \times \nonumber\\
     \!\!\! &{}& \!\!\! \Psi\left(\frac{R}{\left[ \frac{1}{2}\left( \mathbf{n}+\mathbf{p}_{ij} \right)^{\textsf{T}}\mathbf{Q}\left( \mathbf{n}+\mathbf{p}_{ij} \right) \right]^{1/2}}\right)  \nonumber\\
    \!\!\! &{}& \!\!\!  =R^{w}E_{R}(\mathbf{Q},w).
\end{eqnarray}
Let
\begin{equation}\label{Fws}
    F_{w}(\mathbf{Q},s)=\frac{E(\mathbf{Q},s)R^s}{s\left(s-w\right)}.
\end{equation}
Then, we use (\ref{intFws2}) to write
\begin{eqnarray}\label{intFws3}
    \!\!\! &{}& \!\!\!\frac{1}{2\pi \mathbbm{i}}\int_{c-\mathbbm{i}\infty}^{c+\mathbbm{i}\infty}
    F_{w}(\mathbf{Q},s)ds \nonumber\\
    \!\!\! &{}& \!\!\! =\frac{1}{2\pi \mathbbm{i}}\int_{c-\mathbbm{i}\infty}^{c+\mathbbm{i}\infty}
    \left[ \frac{E(\mathbf{Q},s)R^s}{w} \left( \frac{1}{s-w}-\frac{1}{s} \right)
    \right]ds \nonumber\\
    \!\!\! &{}& \!\!\! =\frac{1}{w}\left[ R^{w}E_{R}(\mathbf{Q},w)-E_{R}(\mathbf{Q},0)\right],
\end{eqnarray}
or equivalently
\begin{eqnarray}\label{intFws4}
    &{}&\frac{1}{2\pi \mathbbm{i}}\int_{c-\mathbbm{i}\infty}^{c+\mathbbm{i}\infty}
    F_{w}(\mathbf{Q},s)ds  \nonumber\\
    &{}&~~~= \frac{R^w}{w}\left[
    E_{R}(\mathbf{Q},w)-\frac{1}{R^w}E_{R}(\mathbf{Q},0)\right].
\end{eqnarray}
Next we choose the contour $\alpha_{c,T}$ as in Fig.\ref{fig1}. Let
$\beta$ denote the rectangle with vertices $S_1,S_2,S_3$ and $S_4$
(see Fig.\ref{fig1}). We have
\begin{eqnarray}\label{intFws5}
    &{}& \frac{1}{2\pi \mathbbm{i}}\int_{c-\mathbbm{i}\infty}^{c+\mathbbm{i}\infty}
    F_{w}(\mathbf{Q},s)ds - \frac{1}{2\pi \mathbbm{i}}\int_{\alpha_{c,T}}F_{w}(\mathbf{Q},s)ds \nonumber\\
    &{}& =\frac{1}{2\pi \mathbbm{i}}\int_{\beta}F_{w}(\mathbf{Q},s)ds.
\end{eqnarray}
One can use Cauchy's Residue Theorem to calculate the right side of
(\ref{intFws5}). Refereing to (\ref{Fws}), it is evident that
$F_{w}(\mathbf{Q},s)$ has simple poles at $s=0$ and $s=w$ and may
have a simple pole at $s=3$. Since $E(\mathbf{Q},s)$ is analytic at
$s=0$ and $s=w$, we obtain
\begin{eqnarray}
    \label{ResF0wi} \underset{s=0}{\mathrm{Res}}~F_{w}(\mathbf{Q},s)&=& -w^{-1}E(\mathbf{Q},0), \\
    \label{ResF0wii} \underset{s=w}{\mathrm{Res}}~F_{w}(\mathbf{Q},s)&=& w^{-1}R^{w}E(\mathbf{Q},w).
\end{eqnarray}
To calculate the residue at $s=3$ we use (\ref{Fws}) and
(\ref{ResZ_s3}) to write
\begin{eqnarray}\label{ResF3}
    \underset{s=3}{\mathrm{Res}}~F_{w}(\mathbf{Q},s)&=& \frac{R^3}{3(3-w)}~\underset{s=3}{\mathrm{Res}}~E(\mathbf{Q},s)
    \nonumber \\
    &=& \frac{R^3}{6(3-w)}\sum_{i,j=1}^{N}q_{i}q_{j}\left[ \underset{s=3}{\mathrm{Res}}~Z_{\mathbf{Q}}(s,\mathbf{p}_{ij})
    \right] \nonumber \\
    &=&
    \frac{2^{3/2}\pi
    R^3}{3(3-w)\sqrt{\mathrm{det}\mathbf{Q}}}\sum_{i,j=1}^{N}q_{i}q_{j} \nonumber \\
    &=&
    \frac{2^{3/2}\pi
    R^3}{3(3-w)\sqrt{\mathrm{det}\mathbf{Q}}}\left(\sum_{i=1}^{N}q_{i}\right)^2=0,
\end{eqnarray}
where we used the charge neutrality condition for the unit cell in
the last step. Thus, we have shown that although
$Z_{\mathbf{Q}}(s,\mathbf{p})$ has a simple pole at $s=3$, charge
neutrality implies that residue of $F_{w}(\mathbf{Q},s)$ at $s=3$
vanishes. Therefore, using Cauchy's Residue Theorem and equations
(\ref{ResF0wi}), (\ref{ResF0wii}), and (\ref{ResF3}) we conclude that
\begin{eqnarray}\label{intbeta}
    \frac{1}{2\pi \mathbbm{i}}\int_{\beta}F_{w}(\mathbf{Q},s)ds =
    -w^{-1}E(\mathbf{Q},0)+w^{-1}R^{w}E(\mathbf{Q},w). \nonumber\\
\end{eqnarray}
Substituting (\ref{intFws4}) and (\ref{intbeta}) into
(\ref{intFws5}) results in
\begin{eqnarray}\label{intalpha}
    &{}&\frac{1}{2\pi \mathbbm{i}}\int_{\alpha_{c,T}}F_{w}(\mathbf{Q},s)ds \nonumber\\
    &{}& ~~=\frac{R^w}{w}\left[
    E_{R}(\mathbf{Q},w)-\frac{1}{R^w}E_{R}(\mathbf{Q},0)\right]+w^{-1}E(\mathbf{Q},0)\nonumber\\
    &{}&~~~~~-w^{-1}R^{w}E(\mathbf{Q},w).
\end{eqnarray}
On the other hand, with the aid of (\ref{Es}), (\ref{lemma}), and
(\ref{Fws}) the left side of (\ref{intalpha}) can be written as
\begin{eqnarray}\label{intalpha2}
    &{}&\frac{1}{2\pi \mathbbm{i}}\int_{\alpha_{c,T}}F_{w}(\mathbf{Q},s)ds \nonumber\\
    &{}&=\frac{1}{2}\sum_{i,j=1}^{N}q_{i}q_{j}
    \frac{1}{2\pi
    \mathbbm{i}}\int_{\alpha_{c,T}}\frac{Z_{\mathbf{Q}}(s,\mathbf{p}_{ij})R^s}{s(s-w)}ds=O(R^{\epsilon}),\nonumber\\
\end{eqnarray}
for an arbitrary $\epsilon>0$. Thus, (\ref{intalpha}) becomes
\begin{eqnarray}\label{semiF}
    &{}&E_{R}(\mathbf{Q},w)-\frac{1}{R^w}E_{R}(\mathbf{Q},0)=\nonumber\\ &{}& -\frac{1}{R^w}E(\mathbf{Q},0)+E(\mathbf{Q},w)+R^{-w}O(R^{\epsilon}).
\end{eqnarray}
Since $\mathfrak{Re}(w)>0$, upon taking the limit of (\ref{semiF})
as $R\rightarrow\infty$ and replacing $w$ with $s$, we obtain
(\ref{Es_theorem}). This completes the proof. $~~~~~\square$

\vskip 0.4 in
\section {\textbf{Concluding Remarks}}

A rigorous proof for the Wolf method is given in this paper.
However, there are still some open questions. As we mentioned
earlier, we prove that the undamped Wolf method converges to the
Ewald sum. But usually the undamped method converges very slowly and
this makes it unfavorable in practice. To resolve this issue, Wolf,
et al. \citep{Wolf99} modified their method and introduced the
damped method. The electrostatic energy computed via damped method,
$\mathcal{E}^{D}_{\text{Wolf}}$, is
\begin{equation}\label{DWM}
    \mathcal{E}^{D}_{\text{Wolf}}=\lim_{R\rightarrow\infty}E_{R}^{D}(\mathbf{Q},1),
\end{equation}
where
\begin{eqnarray}\label{EDR}
    &{}&E^{D}_{R}(\mathbf{Q},s)=\frac{1}{2}\sum_{i,j=1}^{N}q_{i}q_{j}
      \sideset{}{'}\sum_{\mathbf{n} \in \mathbb{Z}^3,\mathcal{R}(\mathbf{n},\mathbf{p}_{ij})\leq  R}
    \Bigg\{\nonumber \\ \!\!\!&{}&\!\!\! \frac{ \mathrm{erfc}\left(\alpha \left[ \frac{1}{2}\left( \mathbf{n}+\mathbf{p}_{ij} \right)^{\textsf{T}}\mathbf{Q}\left( \mathbf{n}+\mathbf{p}_{ij} \right)
    \right]^{1/2}\right) }
     {\left[ \frac{1}{2}\left( \mathbf{n}+\mathbf{p}_{ij} \right)^{\textsf{T}}\mathbf{Q}\left( \mathbf{n}+\mathbf{p}_{ij} \right)
    \right]^{s/2}} -\frac{\mathrm{erfc}(\alpha R)} {R} \Bigg\}
    \nonumber \\
    \!\!\!&{}&\!\!\! - \left( \frac{\mathrm{erfc}(\alpha R)}{2R} + \frac{\alpha}{\sqrt{\pi}} \right)
    \sum_{i=1}^{N}q_{i}^{2},
\end{eqnarray}
with
\begin{equation}
    \mathrm{erfc}(x) = 1-\mathrm{erf}(x),~~\mathrm{erf}(x) = \frac{2}{\sqrt{\pi}} \int_{0}^{x}e^{-t^2}dt.
\end{equation}
Since $\underset{x\rightarrow\infty}{\lim}\mathrm{erfc}(x)=0$, using
(\ref{Es_theorem2}) we obtain
\begin{eqnarray}\label{EWolff}
    \!\!\!&{}&\!\!\! \mathcal{E}^{D}_{\text{Wolf}}-\mathcal{E}_{\text{cell}}= -\lim_{R\rightarrow\infty} \Bigg\{
    \frac{1}{2}\sum_{i,j=1}^{N}q_{i}q_{j}\sideset{}{'}\sum_{\mathbf{n} \in \mathbb{Z}^3,\mathcal{R}(\mathbf{n},\mathbf{p}_{ij})\leq  R}
    \Bigg[ \nonumber \\
     \!\!&{}&\!\! \frac{ \mathrm{erf}\left(\alpha \left[ \frac{1}{2}\left( \mathbf{n}+\mathbf{p}_{ij} \right)^{\textsf{T}}\mathbf{Q}\left( \mathbf{n}+\mathbf{p}_{ij} \right)
    \right]^{1/2}\right) }
     {\left[ \frac{1}{2}\left( \mathbf{n}+\mathbf{p}_{ij} \right)^{\textsf{T}}\mathbf{Q}\left( \mathbf{n}+\mathbf{p}_{ij} \right)
    \right]^{1/2}} -\frac{\mathrm{erf}(\alpha R)} {R} \Bigg]
    \Bigg\}
    \nonumber \\ \!\!&{}&\!\! -
    \frac{\alpha}{\sqrt{\pi}}\sum_{i=1}^{N}q_{i}^{2}.
\end{eqnarray}
Although damped method converges fast, it converges to values that
depend on the damping parameter and crystal structure
\citep{Wolf99,YaOrBh2006a}. This means that the right side of
(\ref{EWolff}) converges to values that depend on $\alpha$. Thus, it
does not seem that one can prove a theorem counterpart to one
presented here for the undamped method. But one may find an
interval(s) for the values of the damping parameter to control the
error of the damped method. Also it is interesting to note that the
variation of the calculated energy versus the radius of the charge
natural sphere depends on the crystal structure: it may have an
oscillatory behavior as for NaCl crystal \cite{Wolf99} or
non-oscillatory behavior as for PbTiO$_3$ crystal
\cite{YaOrBh2006a}.

We saw that the charge neutralization idea works for spherical
expanding domains. One may want to see if this idea would work for
other (convex) domains as well. Answer to this question can help to
justify the use of the Wolf method for other geometries like free
surfaces, slabs and regions near crystal defects. Besides
calculating energy, Wolf, et al. \citep{Wolf99} proposed a method
for obtaining forces exerted on charges again without a rigorous
proof. As force is a vector quantity, there is an ambiguity on how
one should project mirror charges on the surface of the sphere.
Since energy only depends on the distance between charges such
ambiguity does not arise in the calculation of energy. Providing
rigorous proofs for the above questions will be the subject of
future work.


\begin{references}

\bibitem{Madelung1918} E. Madelung,
                {\it Physikalische Zeitschrift} {\bf 19} (1918) 524.

\bibitem{El98} E. Elizalde,
            {\it Com. in Math. Phys.} {\bf 198} (1998) 83.

\bibitem{Bor85} D. Borwein, J. M. Borwein, K. F. Taylor,
            {\it J. Math. Phys.} {\bf 26} (1985) 2999.

\bibitem{Bor98} D. Borwein, J. M. Borwein, C. Pinner,
            {\it Tran. American Math. Soc.} {\bf 350} (1998) 3131.

\bibitem{Ewald21} P. P. Ewald,
            {\it Annalen der Physik} {\bf 64} (1921) 253.

\bibitem{Terras85} A. Terras, {\it Harmonic Analysis on Symmetric Spaces and Applications I}
             (Springer-Verlag, Berlin, 1985).

\bibitem[Crandall and Buhler(1987)]{CranBuh87} R. E. Crandall, J. P. Buhler,
            {\it J. Phys. A: Math. and Gen.} {\bf 20} (1987) 5497.

\bibitem{Fu94} K. Fuchizaki,
        {\it J. Phys. Soc. of Japan} {\bf 63} (1994) 4051.

\bibitem{Wolf99} D. P. Wolf, P. Keblinski, S. R. Phillpot, J. Eggebrecht,
              {\it J. Chem. Phys.} {\bf110} (1999) 8254.

\bibitem[Buhler and Crandall(1990)]{BuhCran90} J. P. Buhler, R. E. Crandall,
            {\it J. Phys. A: Math. and Gen.} {\bf 23} (1990) 2523.

\bibitem{Ti86} E. C. Titchmarsh, {\it The Theory of the Riemann Zeta-Function}
             (Oxford University Press, Oxford, 1986).

\bibitem{Ivic85} A. Ivi\'{c}, {\it The Riemann Zeta Function: Theory and Applications}
            (Wiley, New York, 1985).

\bibitem{Murr84} J. D. Murray, {\it Asymptotic Analysis}
             (Springer-Verlag, New York, 1984).

\bibitem{YaOrBh2006a} A. Yavari, M. Ortiz, K. Bhattacharya,
              {\it Philos. Mag.} {\bf 87} (2007) 3997.




\end{references}
\end{document}